\documentclass[prd,aps,twocolumn,superscriptaddress,amsmath,amssymb,nofootinbib]{revtex4-2}
\usepackage{epsfig}
\usepackage{graphicx}
\usepackage{amsmath,amssymb}
\usepackage{dsfont}
\usepackage{bm}
\usepackage{color}
\usepackage[dvipsnames]{xcolor}
\usepackage{upgreek}
\usepackage{mathtools}
\usepackage{bm}
\usepackage{hyperref} 
\usepackage{stmaryrd} 
\usepackage{braket}
\usepackage{physics}
\usepackage{comment}
\usepackage{tcolorbox}
\usepackage{enumerate}
\usepackage{multirow} 

\definecolor{darkblue}{HTML}{001199} 

\definecolor{darkgreen}{HTML}{007700} 

\begin{document}
\title{Emulation of large-scale qubit
registers with a phase-space approach}

\author{Christian de Correc} \email{c.de-correc@bull.com}
\affiliation{Universit\'e Paris-Saclay, CNRS/IN2P3, IJCLab, 91405 Orsay, France}
\affiliation{Eviden Quantum Lab, 78340 Les Clayes-sous-Bois, France}

\author{Denis Lacroix } \email{lacroix@ijclab.in2p3.fr}
\affiliation{Universit\'e Paris-Saclay, CNRS/IN2P3, IJCLab, 91405 Orsay, France}

\author{Corentin Bertrand} \email{corentin.bertrand@bull.com}
\affiliation{Eviden Quantum Lab, 78340 Les Clayes-sous-Bois, France}

\date{\today}
\begin{abstract}
A phase-space approach is used and benchmarked for the simulation of the continuous-time evolution of large registers of qubits.
It is based on a statistical ensemble of independent mean-field trajectories, where mean field is introduced at the level of the qubits, substituting quantum fluctuations/correlations with classical ones. 
The approach only involves at worse a quadratic cost in the system size, allowing to simulate up to several thousands of qubits on a classical computer.
It provides qualitatively accurate description of one-qubit observables' evolutions, making it a useful reference in comparison to techniques limited to small qubit numbers.
The predictive power is, however, less robust for multi qubit observables.
We benchmark the method on the $k$-local transverse-field Ising model, considering a large variety of systems ranging from local to all-to-all interactions, and from weak to strong coupling regimes, with up to 2000 qubits.
To showcase the versatility of the approach, simulations on two-dimensional and three dimensional Ising models are also made.
\end{abstract}


\maketitle

\section{Introduction}

As the number of qubits in quantum computers will increase, and the accuracy in performing quantum operations will improve, quantum processors may out-power classical computer on at least some dedicated tasks.
In particular, this is expected to considerably change our approach to simulating the dynamics of quantum systems.
It raises the question of how to validate the quality of such quantum simulations once we enter the quantum computational era.
In that regard, it will be crucial to be able to provide classically computed comparison points, at least in some specific limits, but for large number of qubits. 

Many powerful classical emulation techniques have been developed to compare with and validate quantum computations.
Tensor network methods form a prominent family in this regard, including Matrix Product States (MPS)~\cite{Orus2014, verstraete-2006-mps-ground-states, schollwock-2011-mps-dmrg, cirac-2021-mps, Verstraete2008}, which are especially good for 1-dimensional systems with low entanglement and local connectivity.
Intensive efforts have also been made to depart from these restrictions, for instance, to describe multi-dimensional systems, with alternative tensor geometries~\cite{Verstraete2004,Verstraete2008, Vidal2007,Vidal2008}.  
Parallel to tensor networks, several families of algorithms have emerged in recent years, such as Pauli truncation techniques based on the stabilizer formalism~\cite{Gottesman1998, Begusic2024, Angrisani2025, rall-2019-pauli-propagation}.
Nevertheless, these are always limited in either the number of qubits, dimensionality, entanglement or magic of the quantum computation.

Our aim in the present work is to benchmark a completely different strategy, based on the phase-space simulation of quantum mechanics~\cite{Gardiner2004}. 
Phase-space approaches aim at replacing a quantum problem by a set of classical-like trajectories~\cite{Wyatt2005,Sanz2012,Sanz2014,Oriols2019}. 
These techniques are semi-classical in nature, but they can eventually incorporate quantum effects depending on their degree of sophistication. 
Among the examples of phase-space techniques providing reformulation of quantum mechanics, one can mention the Nelson stochastic mechanics~\cite{Nelson1966,Nelson1985}, the Bohmian theory~\cite{Bohm1952a, Bohm1952b, deotto-1998-bohmian, durr-2009-bohmian}, the truncated Wigner approximation~\cite{davidson-2017-fTWA, wurtz-2018-cluster-twa, braemer-2024-cluster-twa}, and the Feynman path integrals~\cite{Feynman1948, Feynman1951, Feynman1965, feynman-1979, grosche-1998-feynman, albeverio-2008-feynman}.

We use here an approach inspired by Mori~\cite{Mori1965}, that was first proposed to describe nuclear many-body fermionic systems~\cite{Ayik2008, Lacroix2014}. 
At the heart of our method is a population of random initial states, classically evolved with a mean-field (MF) dynamics, and then averaged to retrieve the more complex quantum dynamics.
The approach, called hereafter Phase-Space Approximation (PSA), was then generalized and applied with success to different fermionic many-body problems in nuclear and condensed matter physics~\cite{Lacroix2012, Lacroix2013, Lacroix2014, Lacroix2014b,Yilmaz2014,Ibrahim2019}. 
Recently, using its early formulation with the spin algebra SU(2) in Ref.~\cite{Lacroix2012}, the PSA was applied to the description of the neutrino oscillation problem using either the SU(2) or SU(3) algebra~\cite{Lacroix2022, Lacroix2024, Kiss2025, Mangin-Brinet2025}. 
These applications, however, correspond to rather specific physical problems involving particles with all-to-all interactions. 
Here we extend its formulation to qubits with varying interactions ranging from nearest-neighbor to all-to-all.

Technically, the PSA on qubits coincides with the discrete truncated Wigner approximation~\cite{schachenmayer-2015-spin-dtwa, kunimi-2021-twa-spin-systems, khasseh-2020-dTWA-Ising, singh-2022-dTWA, shenoy-2024-dtwa-neural-qstates, hosseinabadi-2025-twa-dissipative} (dTWA).
Nevertheless, both approaches originate from different fields and have different formulations.
In this work, we propose a formulation which highlights the role of the density matrix, and which we believe is more adapted to the quantum computing community.
By comparison with previous benchmarks~\cite{schachenmayer-2015-spin-dtwa, kunimi-2021-twa-spin-systems, shenoy-2024-dtwa-neural-qstates}, we provide a more comprehensive study in interaction strength, interaction range, and initial state.
We also put an emphasis on the dynamics of large systems, with a comparison to MPS evolution in systems with up to 2000 qubits.

The present work has two main objectives: (i) to provide a simple formulation of the PSA/dTWA technique for qubits; (ii) to systematically assess the predictive power of the approach for a wide class of problems and clarify the type of observables that could be safely predicted with this method. 
To this aim, we benchmark the technique using the $k$-local Transverse Field Ising Model (TFIM) encoded on qubits with $k$ ranging from $1$ (nearest neighbor interaction) to $k=L-1$, where $L$ is the number of qubits. 
The latter case corresponds to an all-to-all interaction between qubits, a case similar to the Lipkin-Meshkov-Glick model \cite{Lip65a,Mes65,Gli65} that is often used in nuclear physics to benchmark many-body approaches~\cite{Hol73,Rob92,Sev06}.

This article is organized as follows.
In Sec.~\ref{sec:TFIM-and-MF}, we introduce the 1D TFIM and its mean-field treatment, setting notations for the benchmark.
Mean field is a stepping stone to the description of the PSA, which we complete in Sec.~\ref{sec:psa}. The results of our benchmark on small scale problems are detailed in Sec.~\ref{sec:psa-benchmark-small-scale}.
Finally, results on large scale problems are shown in Sec.~\ref{sec:psa-benchmark-large-scale}, where we also illustrate our technique on high-dimensional geometries.

\section{The $k$-local TFIM and its mean-field approximation}
\label{sec:TFIM-and-MF}

We consider the 1D TFIM with open boundary conditions, and with interaction between neighbor spins at a distance of at most $k$. 
For $L$ sites, the Hamiltonian takes the form:
\begin{equation}
    \label{eq:hamTFIM}
    H = -h \sum_{i = 1}^L Z_i 
    - J\sum_{i = 1}^L
    \sum_{j \sim i}
    X_i X_j, 
\end{equation}
with $(X_i, Y_i, Z_i)$ the Pauli matrix acting on the qubit $i$, and the relation $\sim$ describes the $k$-local connectivity of the chain:
\begin{equation}
    j \sim i 
    \quad
    \Longleftrightarrow
    \quad
    \left(
    j > i
    , \quad
    \vert j - i\vert \le k \right) .
\end{equation}
We refer to this as the $k$-local TFIM model. 
To simplify, we will set $\hbar = 1$ and take $h$ as the unit of energy.

\subsection{Energy landscape of the mean-field eigenstates}

As a first step, we want to expose the physics of the model and introduce the relevant parameters for our benchmark.
A mean-field approach is convenient to reveal the well-known paramagnetic-to-ferromagnetic quantum phase transition when increasing the coupling strength $J$. 
To see this, consider an arbitrary translation-invariant product state $\ket{\theta, \phi}^L$, where
\begin{equation}
    \label{eq:def-theta-state-1-qubit}
    \ket{\theta, \phi} 
    = 
    \cos(\theta / 2) \ket{0} + e^{i\phi}\sin(\theta / 2)\ket{1},
\end{equation}
is any single qubit state described by $0 \le \theta \le \pi$ and $0 \le \phi \le 2\pi$.
This satisfies
\begin{equation}
    \expval{X_i} 
    = 
    \cos\phi \sin\theta,
    \qq{and}
    \expval{Z_i} 
    = 
    \cos\theta.
\end{equation}
Irrespective of $k$ (and of the boundary conditions or dimensionality of the problem), the total energy of this ansatz can be expressed as
\begin{equation}
    \label{eq:mf-ansatz-energy}
    E(\theta, \phi) 
    =
    -hL \left( \cos\theta 
    + {\eta \over 2} \cos^2\phi \sin^2\theta \right), 
\end{equation}
where the parameter $\eta$ is a function of $h$, $J$, $L$ and of the connectivity of the Hamiltonian:
\begin{equation}
\label{eq:eta-J}
    \eta 
    = 
    \frac{2J}{h} \left( k - {k(k+1) \over 2L} \right).
\end{equation}
Since $k \le L-1$, we see that $\eta \ge 0$, so that the energy is always minimal for $\phi=0$.

\begin{figure}[htbp]
    \centering
    \includegraphics[width=1\linewidth]{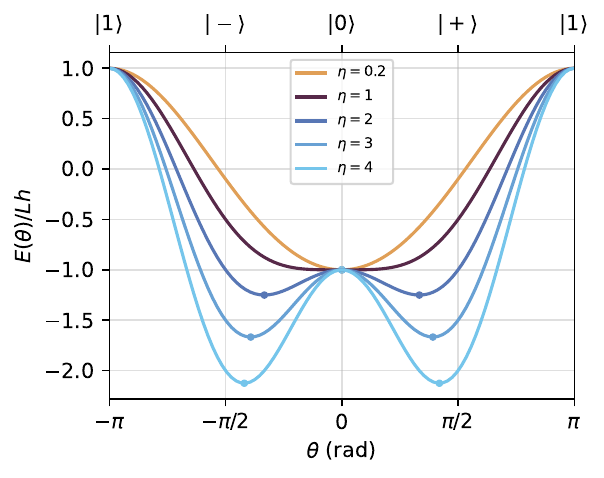}
    \caption{Mean-field eigenenergies of the TFIM model given by Eq. (\ref{eq:mf-ansatz-energy}) for different values of $\eta$, in a chain of length $L = 12$.
    Thicker points indicate energy minima.
    }
    \label{fig:angular-dependance-mean-field}
\end{figure}

Depending on whether $\eta \le 1$ or not, the energy $E(\theta) = E(\theta, \phi=0)$ admits one or two minima for $\pm \theta_\textrm{min}$ located in the region $(-\pi/2, \; \pi/2)$.
This is illustrated in Fig.~\ref{fig:angular-dependance-mean-field} for different values of $\eta$, where a
transition occurs from a single minimum of $E(\theta)$ at $\theta_\textrm{min}=0$ ($\eta < 1$), to two minima located at some $\pm \theta_\textrm{min} \neq 0$.
This phase transition is associated to the parity symmetry $X_i \rightarrow -X_i$, which is respected for $\eta < 1$ (paramagnetic phase, $\langle X_i \rangle_{\theta_\textrm{min}} = 0$) and broken for $\eta > 1$ (ferromagnetic phase, $\langle X_i \rangle_{\pm\theta_\textrm{min}} \neq 0$).
Since $\eta$ decreases with $k$, increasing connectivity favors the paramagnetic phase.

In order to compare simulations with different values of $k$, we will take $\eta$ as the main parameter of our model instead of $J/h$.

\begin{figure*}[!htbp]
    \centering
    \includegraphics[width=1\linewidth]{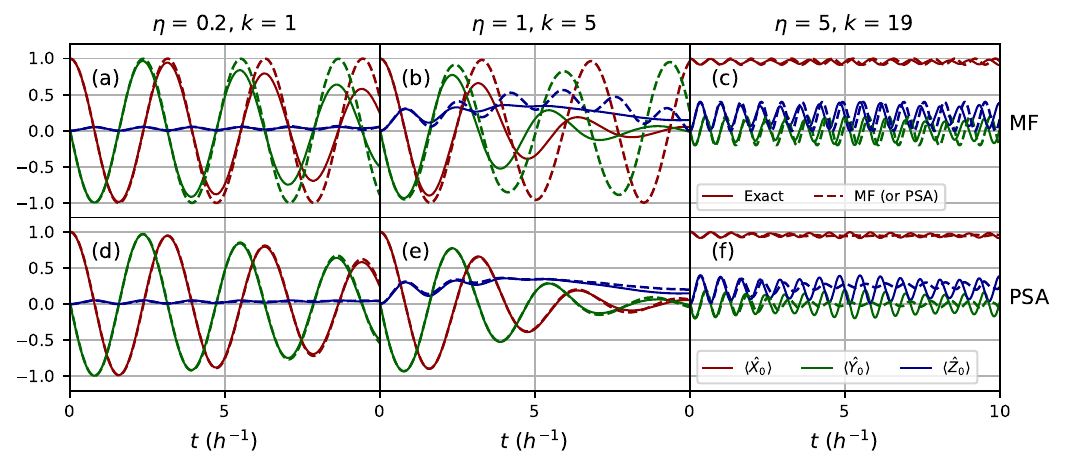}
    \caption{
    Comparison of the MF (dashed, top row) and PSA (dashed, bottom row) to the exact (plain, both rows) evolutions of Pauli matrices expectation values, with an advantage of the PSA over the MF. 
    Specifically, the evolution of the observables $\hat{S}_O = {1\over L}\sum_i \hat{O}_i$, with $\hat{O} = \hat{X}, \hat{Y}, \hat{Z}$, is shown as a function of time for three different regimes of ($\eta$, $k$), using the initial state $\ket{+}^L$ in a chain of length $L = 20$. 
    }
    \label{fig:exa-psa-mf-comparison}
\end{figure*}

\subsection{Mean-field dynamics}
\label{sec:mfstrategy}

Mean-field (MF) theory provides a simple approximate method to study the evolution of a system.
It is both a theoretical inspiration for PSA and a relevant method to compare with, which justifies our detailed introduction here.
It has been heavily studied, in particular in the Ising model~\cite{strecka-2015-tfim-MF, negele-1982-MF, jensen-1983-ising-mf, zimmer-1993-ising-mf, waldorp-2025-ising-mf, wang-2012-ising-mf, grest-1986-ising-mf, punya-2010-ising-mf}.

A simple and intuitive way to introduce MF theory is to consider the evolution of an operator $O$ dictated by Ehrenfest's equation:
\begin{equation}
    \label{eq:ehrenfest-theorem}
    i \frac{d\langle O\rangle}{dt}
    =
    \expval{
        [O, H]
    }.
\end{equation}
Solving exactly all the Ehrenfest equations is expected to be as hard as solving the Schrödinger equation.
This resolution is expensive since the exponential dimension of the underlying Hilbert space induces an exponential computational cost.
In the case of the Ehrenfest equations, this manifests as the coupling induced by the $\expval{[O, H]}$ terms between exponentially many different observables.
Thus, solving the Ehrenfest equations for a small number of operators (e.g. local observables) requires either an additional approximation or a prohibitive computational cost.

To circumvent this issue, the MF approach focuses on the three Pauli observables $X_i, Y_i, Z_i$ for each qubit $i$, which form a complete set of observables in the qubit operator space.
Each such operator $O_i$ is associated to a time-dependent scalar $o_i$, set at initial time $t_0$ to $o_i(t_0) = \expval{O_i}(t_0)$.
The MF approximation then consists in replacing $\expval{O_i}$ by $o_i$ in the Ehrenfest equation, with the convention that on the right-hand side, each Pauli product $\expval{O_1 \dots, O_L}$ in the commutator term of Eq.~\eqref{eq:ehrenfest-theorem} is replaced as follows:
\begin{equation}
    \label{eq:mean-field-approx}
    \expval{O_1 \otimes \dots \otimes O_L}
    \longrightarrow
    o_1
    \times \dots \times 
    o_L.
\end{equation}
Within this approach, where quantum correlations are neglected, the state is approximated by a product state and we expect that
\begin{equation}
    \forall t > 0, 
    \quad
    o_i(t) \simeq \expval{O_i}(t).
\end{equation}
This approximation is exact when the quantum state is a product state, and it remains accurate as long as little entanglement builds up during the evolution, especially in the short term.
Crucially, this MF approximation closes the system of Ehrenfest equations with the $3L$ single-qubit Pauli operators, making its resolution tractable using any ordinary differential equations solver.

In the case of our $k$-local TFIM, this results in this set of equations:
\begin{equation}
\label{eq:mean-field-system-for-TFIM}
\begin{cases}
    \partial_t x_i 
    & = \enspace
    2 h y_i
    \\
    \\
    \partial_t y_i 
    & = \enspace 
    \displaystyle
    - 2 hx_i   
    + 2 J z_i 
    \sum_{j \sim i} 
    x_j
    \vspace{-0.3cm} \\
    \\
    \partial_t z_i 
    & = \enspace
    \displaystyle
    -2 J  y_i  
    \sum_{j \sim i} 
    x_j
\end{cases},
\end{equation}
for all $i$.

By construction, the MF approximation assumes that single-qubit observables contain most of the relevant information to reproduce the system properties, and that each qubit $i$ is described by its time-dependent Bloch vector 
\begin{equation}
    \label{eq:bloch-vector-definition}
    \vec{r}_i 
    = 
    \begin{pmatrix}
     x_i \\
     y_i \\
     z_i   
    \end{pmatrix}
    \simeq
    \begin{pmatrix}
     \expval{X_i} \\
     \expval{Y_i} \\
     \expval{Z_i}   
    \end{pmatrix}
    =
    \vec{r}_i^{\, \textrm{exact}}
    \in \mathbb{R}^3, 
\end{equation}
where the approximation between the two vectors indicates how well the mean-field approximation accurately reproduces the exact evolution. 

The Bloch vector $\vec{r}_i$ can also be viewed as a representation of the 1-qubit reduced density matrix $R_i$, through
the mapping~\cite{Nielsen2010}
\begin{equation}
    \label{eq:mf-1qubit-density-matrix}
    R_i
    =
    {
        1 + \vec{r}_i \cdot \vec{\sigma}
        \over 2 
    }
    =
    {1 \over 2}
    \begin{pmatrix}
        1 + z_i & x_i - iy_i \\
        x_i + iy_i & 1 - z_i 
    \end{pmatrix},
\end{equation}
where $\vec{\sigma}$ is the vector of Pauli matrices.
In the exact solution, the one-qubit density matrix can also be obtained from the full density matrix
by performing the partial trace over all other qubits.  
Conversely, in the MF case, the full density matrix is the tensor product of each single-qubit reduced density matrix, since correlations between qubits are neglected:
\begin{equation}
    \label{eq:mf-full-density-matrix}
    R = R_1 \otimes \dots \otimes R_L.
\end{equation} 
The MF approximation is thus expected to be applicable when the system is always close to a product state, i.e., when quantum fluctuations, or equivalently entanglement, are small all along the time evolution.

To illustrate this, we solved the system of Eq.~\eqref{eq:mean-field-system-for-TFIM} with a 4-th order Runge-Kutta technique, over a time interval $[0, T]$ with $T = 10 \; (h^{-1})$, with a time step $dt = 10^{-2} \; (h^{-1})$.
In Fig.~\ref{fig:exa-psa-mf-comparison} (top row), we compare the MF evolution of $x_i(t)$, $y_i(t)$ and $z_i(t)$ with the exact dynamics of $\expval{X_i}$, $\expval{Y_i}$ and $\expval{Z_i}$.
All these quantities are averaged over the $L$ qubits to make the comparison more legible, a choice justified by the small variance between different qubits (see App.~\ref{app:inter-qubit-variance}).

The results in a chain of length $L = 20$ are displayed for several values of $\eta$ and $k$, with the initial tensor product state $\ket{+}^L$. 
As expected from the crudeness of the MF approximation, we observe that the MF approach can only reproduce the dynamics in a fairly restricted range of parameters.
It succeeds in particular in the case where $J \propto \eta \ll 1$ and $k \gg 1$, for which the coupling terms in Eq.~\eqref{eq:mean-field-system-for-TFIM} vanish, making the evolution perturbative and the MF product state close to exact.
Conversely, we see that in the strong coupling regime $\eta \gg 1$, the MF approach can also be accurate.
This success is due to the fact that here the initial state is almost an eigenstate of $H$ when $J L\propto L\eta \gg h$, making the evolution perturbative. 

\begin{figure}
    \centering
    \includegraphics[width=1\linewidth]{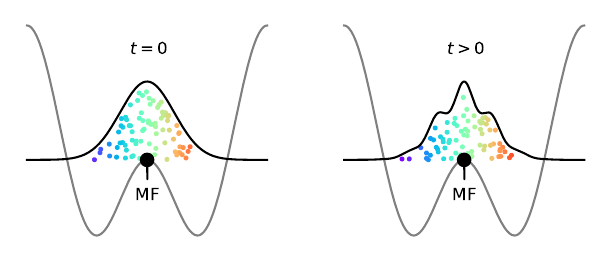}
    \caption{
    Schematic representation of the PSA approach as the average of multiple stochastic MF trajectories, enabling better reproduction of quantum-state properties and subsequent evolution. 
    Each point corresponds to a trajectory, while different colors guide the eyes. Using the mean-field theory for the potential energy surface shown would lead to no evolution (in this specific example where the initial state is $\ket{0}$). 
    Indeed, in an MF approach (black filled circle), one cannot spontaneously break the symmetry of the problem, which would mean here that the system decides to move to the left or right side.
    }
    \label{fig:psa-schema}
\end{figure}

\section{Phase-Space approximation (PSA) for qubits}
\label{sec:psa}

\subsection{General description}

To overcome the limitations of MF, the PSA assumes each $\vec{r}_i$ (or equivalently $R_i$) to be a random variable, characterized by its random initial value at $t = t_0$ and evolved for $t > t_0$ with the MF equations of motion.
Each sample of the initial values produces a so-called trajectory, evolved independently from the others with the MF equations of motion.
Each trajectory is therefore represented by a separable density matrix as in Eq.~\eqref{eq:mf-full-density-matrix}.
In the following, we denote quantities from the $n$-th trajectory (or realization) by an exponent~$\dots^{(n)}$, and statistical averages (means over all trajectories) 
with an overline to distinguish them from quantum expectation values.

To produce the correct initial time observables, the initial sampling is made such that the average (over trajectories) density matrix corresponds to the exact initial state $\rho(t_0)$,
\begin{equation}
	\label{eq:stat-mean-equals-exact-at-t0}
	\overline{R}(t_0)
	=
	{1 \over N_\textrm{traj}}
	\sum_{n = 1}^{N_\textrm{traj}}R^{(n)}(t_0)
	\xrightarrow[N_\textrm{traj} \to +\infty]{}
	\rho(t_0).
\end{equation}
Other conditions are necessary to reproduce the correct quantum fluctuations; they are discussed further below, along with the sampling method we use.
Importantly, the initial time density matrix of a trajectory is not a physical one.
In particular, it lies beyond the Bloch sphere.
Physicality is recovered after averaging over trajectories.

Each $R^{(n)}$ is then evolved within the MF approximation~\eqref{eq:mean-field-approx}, so that $\overline{R}(t)$ provides an approximation of $\rho(t)$ at any $t$.
Importantly, since trajectories are independent, the computational complexity of the method is only $O(N_\textrm{traj} L^2)$, and it easily benefits form multiprocessing computation, making it barely more expensive than the mean-field approach itself.
Furthermore, unlike the bare MF approach, the density matrix $\overline{R}$ can host correlations between qubits, due to correlations between the random variables $R_i$ at different $i$.

Quantum expectation values during the time evolution are replaced by statistical averages over the set of $(x_i, y_i, z_i)$ variables during the evolution. 
For instance, if we want to compute the expectation value of some Pauli string like $X_1 \dots Y_L$, we follow the correspondence:
\begin{align}
    \label{eq:recovering-PSA-expectation-value}
    \langle X_1 \dots Y_L\rangle_t 
    & \longrightarrow {\rm Tr} \left( X_1 \dots Y_L  \overline{R}(t) \right), \nonumber \\
    & = 
    \overline{{\rm Tr} \left[ X_1 R_1(t)\right] ,
    \dots 
    {\rm Tr} \left[ Y_L R_L(t) \right] } , \nonumber \\
    & = 
    \overline{x_1(t) \dots  y_L(t)}.
\end{align}
Importantly, the last line is in general not equal to $\overline{x_1(t)} \times \dots \times \overline{y_L(t)}$  due to the coupling between the different qubits in Eq.~\eqref{eq:mean-field-system-for-TFIM}.
This expression is very useful in practice, since one can obtain the evolution of observables simply by performing a statistical average over their MF scalar equivalent along the trajectory. 
Notably, to obtain observables, it is not necessary to explicitly construct the average density matrix, and the required memory usage remains linear in $L$.

To highlight the PSA philosophy, we give in Fig.~\ref{fig:psa-schema} a schematic illustration of the sampling technique, and the evolution of events expected in the PSA technique for one of the energy landscapes shown in Fig.~\ref{fig:angular-dependance-mean-field}. 

\subsection{Initial sampling}

The PSA relies on a choice of distribution for the initial density matrix $R(t_0)$, so that it satisfies Eq.~\eqref{eq:stat-mean-equals-exact-at-t0}.
This ensures that every observable is recovered exactly at the initial time.
In addition, we need to add constraints to enforce that the correct quantum fluctuations are also reproduced.
Such sampling methods have been discussed in Refs.~\cite{Lacroix2012,Yilmaz2014,Ibrahim2019}, with different sets of constraints and solutions.

Here, we assume that the initial state is a product state.
As a result, we treat the different $\vec r_i(t_0)$ as statistically independent variables, and focus on the probability distribution to sample each coordinate for the qubit $i$. 
At this step, we highlight that full quantum correlations cannot be recovered by classical probabilities in the PSA framework, due to the fact that $X$, $Y$, and $Z$ do not commute.
For instance, despite the relation $\expval{X_iY_i} = i\expval{Z_i}$, it is not possible to satisfy the constraint $\overline{x_i} \times  \overline{y_i} = \overline{x_i y_i} = i \, \overline{z_i}$
due to the variables $\overline{x_i}, \overline{y_i}, \overline{z_i}$ being real-valued.

Despite our model being unable to reproduce such relations on products of different observables, we can impose that all moments of the expectation value of $X_i$, $Y_i$, $Z_i$ are reproduced by the moments of the random variables $x_i$, $y_i$, $z_i$.
Formally, at any order $m \ge 1$ and at $t=t_0$, this condition is rephrased as
\begin{equation}
    \label{eq:equality-higher-moments-at-t0}
    \overline{x_i^m} 
    =
    \langle X_i^m \rangle,
    \quad
    \overline{y_i^m} 
    =
    \langle Y_i^m \rangle,
    \quad
    \textrm{and}
    \quad
    \overline{z_i^m} 
    = 
    \langle Z_i^m \rangle.
\end{equation}
As it turns out, despite Eq.~\eqref{eq:equality-higher-moments-at-t0} being a stronger assumption than Eq.~\eqref{eq:stat-mean-equals-exact-at-t0} (which is equivalent to the case $m=1$), it is easily satisfied. To do so, we take a biased Rademacher distribution~\cite{rademacher-1922} with support in $\{-1, +1 \}$, defined by:
\begin{equation}
    \begin{cases}
        z_i(t_0) = + 1 \\
        z_i(t_0) = - 1
    \end{cases}
    \qq{with probability}
    \begin{cases}
        \bra{0}\rho_i(t_0)\ket{0} \\
        \bra{1}\rho_i(t_0)\ket{1}.
    \end{cases}
\end{equation}
and similarly for $x_i, y_i$, using the eigenbasis of $X_i$ and $Y_i$ instead of $(\ket{0}, \ket{1})$.
This corresponds to the probability distribution defined by Born's rule, assuming the state is prepared anew between each measurement.
Due to the fact that $x_i^2 = y_i^2 = z_i^2 = 1$, odd (respectively even) moments are all equal (same as for Pauli observables), and it is easily seen that the correct moments are reproduced for $m=1,2$ by Born's rule.

To sum up, the distribution of initial density matrices is obtained by sampling the Bloch coordinates.
Each coordinate is sampled independently, for lack of a better way to reproduce quantum fluctuations with classical probabilities.
Furthermore, for each coordinate, a natural choice of sampling is to copy Born's rule probability law, ensuring the moments at any order $m \ge 1$ associated with the exact state $\rho(t_0)$ are well reproduced.
We conclude by highlighting the following fact:  each qubit having a Bloch norm $\Vert \vec{r}_i(t_0) \Vert = \sqrt3$, which is conserved along the MF evolution, the trajectories themselves are not physical.
This is however not problematic, since we only interpret the average over the trajectories, which satisfies $\Vert \vec{r}_i(t) \Vert \le 1$ up to some statistical fluctuations.


\subsection{Discussion of the method for a general initial state}

The PSA can also be formulated for complex and correlated initial states~\cite{Yilmaz2014}. 
In this case, the condition that the average density matrix matches the exact density matrix can be relaxed. 
Instead, a minimal condition should be that the expectation values of Pauli strings of length one and two are properly reproduced. 
This is equivalent to imposing that the first and second moments of 1-qubit observables are properly described at the initial time, as was originally proposed in Ref.~\cite{Ayik2008}.

Effects beyond MF are incorporated through the statistical average and through the fact that the MF equations of motion, or Eqs.~\eqref{eq:mean-field-system-for-TFIM}, are non-linear. 
Such equations also highlight the importance of reproducing the first and second moments at $t_0$. 
Indeed, by averaging the Eqs.~\eqref{eq:mean-field-system-for-TFIM} over the trajectories, 
one observes that $(\overline{x_i} ,\overline{y_i}, \overline{z_i}$ are coupled to the set of second moments $\{ \overline{z_i x_j},\overline{y_i x_j} \}$, similarly to what we have in the exact solution. 
More generally, one can demonstrate that the PSA leads to a set of coupled moments where the $k$-th moments are coupled to the $(k+1)$-th moments, a feature that significantly enriches the dynamics~\cite{Lacroix2016}.

\section{Benchmark on small-scale TFIM}
\label{sec:psa-benchmark-small-scale}

To assess the predictive power of the PSA, we have systematically compared the evolution of observables mean values for varying two-body strength $\eta$, connectivity $k$, and for different initial states.
In the following, all PSA results will be obtained by averaging over $N_\textrm{traj} = 10^4$ trajectories, which is sufficient to achieve convergence.

\subsection{Single-qubit observables}

First, we show in Fig.~\ref{fig:exa-psa-mf-comparison} (bottom row) several examples of evolution, with the same parameters as for the MF results (top row). 
By comparison with the bare MF approach, Fig.~\ref{fig:exa-psa-mf-comparison} illustrates how the PSA technique has a generally better predictive power to simulate 1-qubit averaged observables. 
In particular, it reproduces the damping mechanism and the long-time asymptotics over a wide range of $k$ and $\eta$ values. 
Similar improvements were observed in its original many-body formulation~\cite{Lacroix2012,Lacroix2013,Lacroix2014,Yilmaz2014} or when applied to neutrino systems~\cite{Lacroix2022,Lacroix2024,Mangin-Brinet2025}. 
Above $\eta = 2$, the system enters a different regime for large $k$ values, where the two-qubit interaction becomes dominant and the one-qubit term acts as a perturbation. 
Surprisingly enough, MF turns out to outperform the PSA where the damping is overestimated while the exact solution continues to oscillate; see also App.~\ref{app:systematic-scanning} for more results in this regime. 
Such behavior is nonetheless rather specific to the initial state we considered, $\ket{+}^{L}$, which is an eigenstate of the two-qubit terms of the Hamiltonian. 

To assess the dependency in the initial state, we consider initial states of the form $\ket{\theta, \phi}^L$ with the definition in Eq.~\eqref{eq:def-theta-state-1-qubit}.
The deviation of the MF or PSA approach, with respect to the exact dynamics, is then quantified by the following time- and qubit-averaged error:
\begin{equation}
    D_r(T) 
    =
    \frac{1}{LT} \int_0^{T} 
    \sum_{i=1}^L 
    \Vert \vec{r}_i^{\textrm{exact}}(t) - \vec{r}_i(t) \Vert
    \textrm{d}t
\end{equation}
with the time-dependent Bloch vectors of Eq.~\eqref{eq:bloch-vector-definition}.
Here $\Vert \cdot \Vert $ is the $L^1$-norm, and $T$ is the total simulation time.

\begin{figure}[!ht]
    \centering
    \includegraphics[width=1\linewidth]{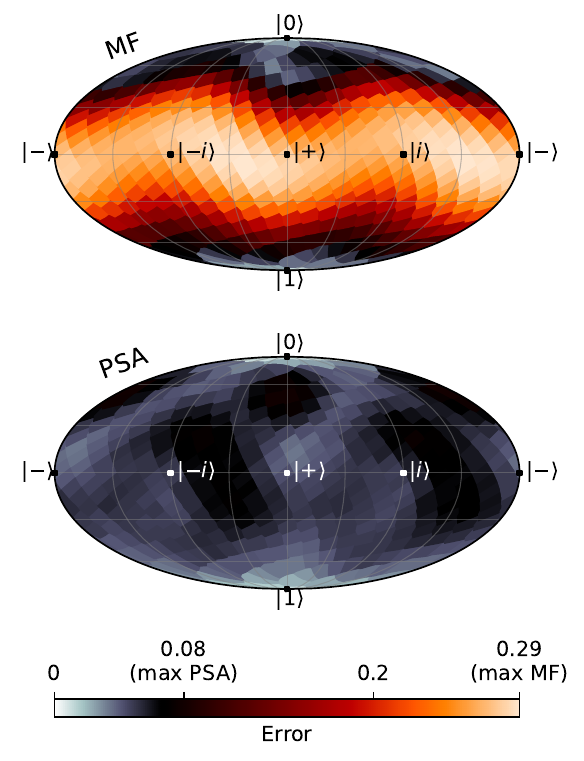}
    \caption{
    Deviation $D_r(T)$ as a function of the initial state $\ket{\theta, \phi}^L$, shown on the Bloch sphere.
    For most initial states, the PSA outperforms the MF, apart near the two poles $\ket{0}$ and $\ket{1}$.
    Parameters: $k = 1$, $\eta = 0.5$, $T = 10h^{-1}$.
    }
    \label{fig:psa-heatmap-mollweide}
\end{figure}

We compute this quantity for 400 different $(\theta, \phi)$ angles sampled from a near-uniform distribution on the Bloch sphere (see App~\ref{app:influence-initial-state} for more details), using $k = 1$, $\eta = 0.5$.
The resulting values of $D_r(T)$ are shown in Fig.~\ref{fig:psa-heatmap-mollweide} for both the MF and PSA approaches, where each $(\theta, \phi)$ is located at the place of $\ket{\theta, \phi}$ on the Bloch sphere.
As a direct consequence, the PSA appears to be much more robust than the MF with respect to the choice of initial state.

\begin{figure*}[!ht]
    \centering
    \includegraphics[width=1\linewidth]{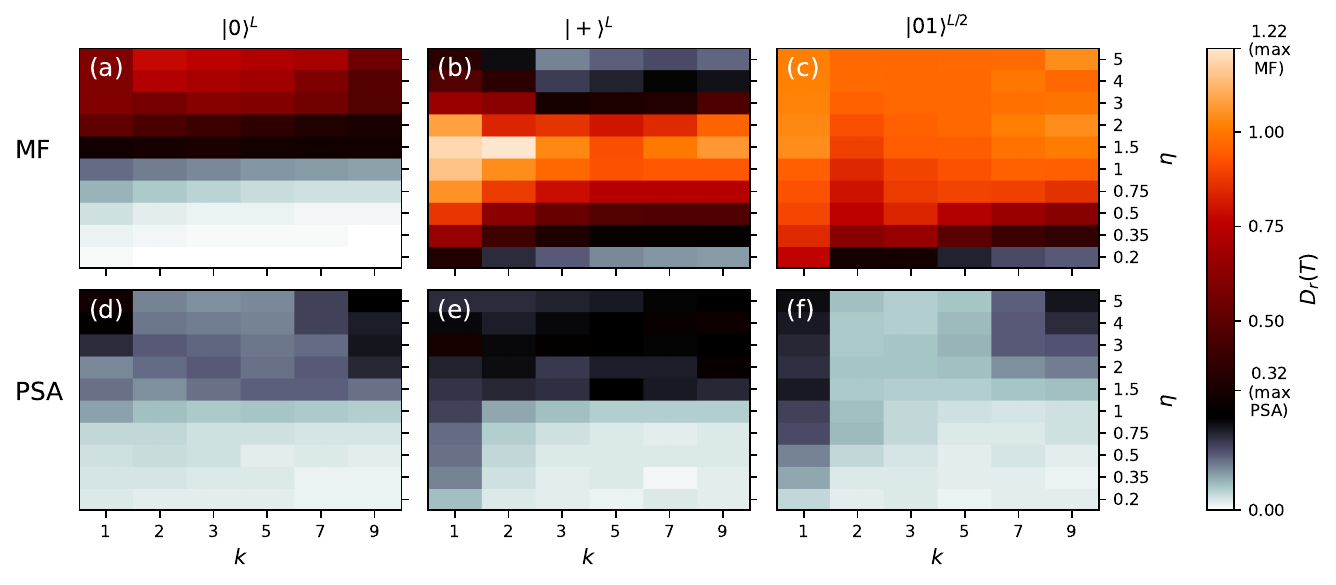}
    \caption{
    Deviation from the exact dynamics $D_r(T)$, for different $(k,\eta)$ values. 
    From left to right, the initial state is $|0\rangle^L$, $\vert + \rangle^L$ and $\vert 01 \rangle^{L/2}$, and MF (top row) is compared to PSA (bottom row). 
    Results have been obtained for $L=10$ and final time $T=10 \; (h^{-1})$.
    }
    \label{fig:psa-heatmap}
\end{figure*}

The PSA approach improves upon MF by inducing a much better reconstruction of the Bloch vector, and this advantage is most important near the equator of the Bloch sphere.
Interestingly, very close to the poles $\ket{0}$ and $\ket{1}$, both techniques display similar performances; this phenomenon is attributable to the fact $\ket{0}^L$ and $\ket{1}^L$ are almost eigenstates of $\hat{H}$ in the paramagnetic regime ($\eta < 1$).

Then, we compute the deviation $D_r(T)$ along a grid of the $k$ and $\eta$ parameters, for three different initial state.
The resulting heatmaps are shown in Fig.~\ref{fig:psa-heatmap} as a function of $k$ and $\eta$.

Comparing PSA (bottom row) to MF (top row), we observe a general advantage of the PSA, with once again a high dependence in the $\eta$, $k$ and initial state.
In good agreement with the evolutions displayed in Fig.~\ref{fig:exa-psa-mf-comparison} and in App.~\ref{app:systematic-scanning}, a clear transition around the MF critical threshold $\eta = 1$ appears when the input state is $\ket{0}^L$ or $\ket{+}^L$.
While it lacks this transition, the third initial state gives evidence that the observed PSA advantage generalizes for inital states with a lesser degree of symmetry.
Finally, we see in this figure that the PSA typically improves for intermediate values of $k$ (panels (d) and (f) especially), and it degrades for $\eta \ge 1$, where the approximation of independent trajectories breaks down (panels (d) and (e)).

\subsection{Two-qubits observables}

\begin{figure*}[!ht]
    \centering
    \includegraphics[width=1\linewidth]{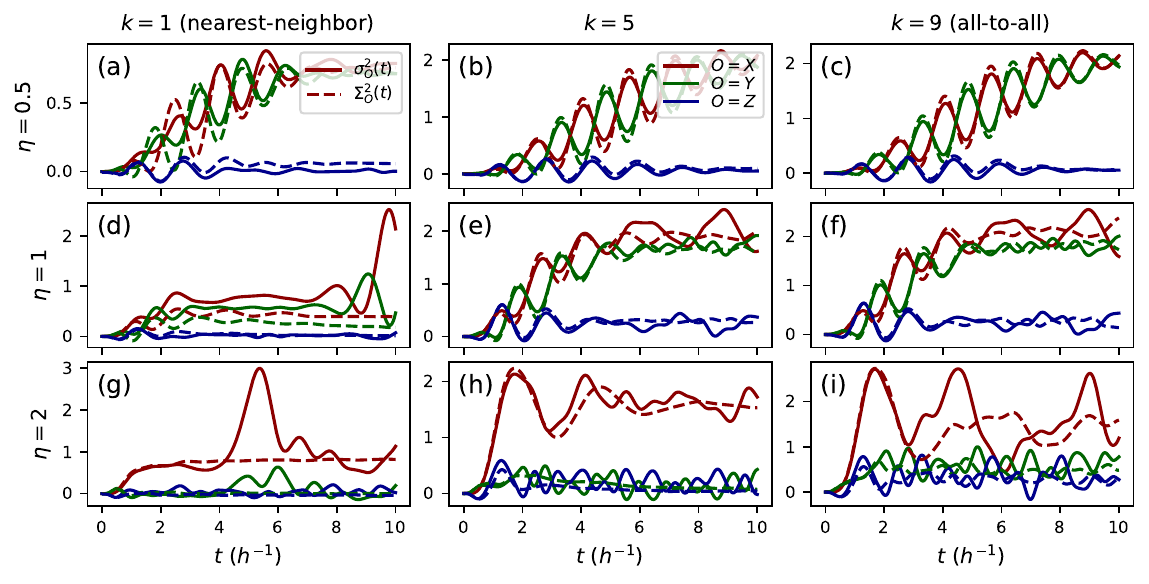}
    \caption{Comparison of the exact correlations $\sigma_O$ (plain lines) with the PSA correlations $\Sigma_O$ (dashed lines) for the three operators $O = X$, $Y$ and $Z$.
    The initial state $\ket{+}^L$ is evolved with the parameters $\eta \in \{ 0.5, 1, 2 \}$ and $k \in \{1, L/2, L-1 \}$ in a chain of length $L = 10$.
    } 
    \label{fig:exa-psa-comparison-fluc}
\end{figure*}

Besides Bloch coordinates, the PSA technique can predict correlations (or fluctuations) and higher-order moments.
We illustrate this aspect in the present section by defining the three quantities:
\begin{equation}
    \label{eq:def-sigma_o}
    \sigma^2_O (t) 
    =
    {1 \over L} 
    \sum_{i < j} 
    \langle O_i O_j \rangle_t - \langle O_i \rangle_t \langle O_j \rangle_t 
\end{equation}
for $O=X$, $Y$ or $Z$. 
Illustrations of the quantum fluctuations of these quantities are shown 
in Fig.~\ref{fig:exa-psa-comparison-fluc}, and compared to their PSA equivalent obtained from the average:

\begin{equation}
    \Sigma^2_O (t) 
    = 
    {1 \over L} 
    \sum_{i < j} 
    \overline{o_i o_j}  - \overline{o_i} \cdot \overline{o_j}
    .
    \label{eq:fluc-oo-psa}
\end{equation}
This figure shows the different parameter regimes in which the PSA succeeds or fails at predicting the exact quantum fluctuations.
A good reproduction is observed when $k > 1$ and $\eta \le 1$, in agreement with the previous analysis made for 1-body observables and synthesized in Fig.~\ref{fig:psa-heatmap}.
For $\eta > 1$, we see faster oscillations that the PSA fails to reproduce.
We observe these oscillations are not well reproduced because the PSA introduces a stronger dampening than the exact. 
This dampening can be attributed to the statistical averaging of many trajectories with slightly different phases, and was seen in other similar works~\cite{braemer-2024-cluster-twa}.
Finally, for $k = 1$ and $\eta = 2$, we observe in Fig.~\ref{fig:exa-psa-comparison-fluc}~(panel g) a high peak of the value of $\sigma_x^2(t)$.
This singular behavior can be attributed to the boundary effect, in which part of the wave function bounces back from the end of the chain, leading to interference.

\subsection{Entanglement entropies}

As a reminder, the PSA framework allows us to recover the time-dependent reduced density matrix $\overline{R}_i$ of qubit $i$ through Eq.~\eqref{eq:mf-1qubit-density-matrix}.
By tensor product, this generalizes to the reduced density matrix $\overline{R}_I = \overline{R_{i_1} \otimes \ldots \otimes R_{i_m}}$ of any subsystem $I = (i_1, \dots, i_m)$ of $m$ qubits.
Given such a subsystem, we are interested in its entanglement entropy with the other qubits,
\begin{equation}
    \label{eq:multi-qubit-entanglement-entropy}
    S_I = - \Tr[\overline{R}_I \log \overline{R}_I].
\end{equation}
If $\overline{R}_I$ is a good approximation to the exact reduced density matrix on $I$, then the entanglement entropy should be reproduced faithfully.
Therefore, this provides a key quantity to evaluate the predictive power of the PSA.

\begin{figure*}
    \centering
    \includegraphics[width=1\linewidth]{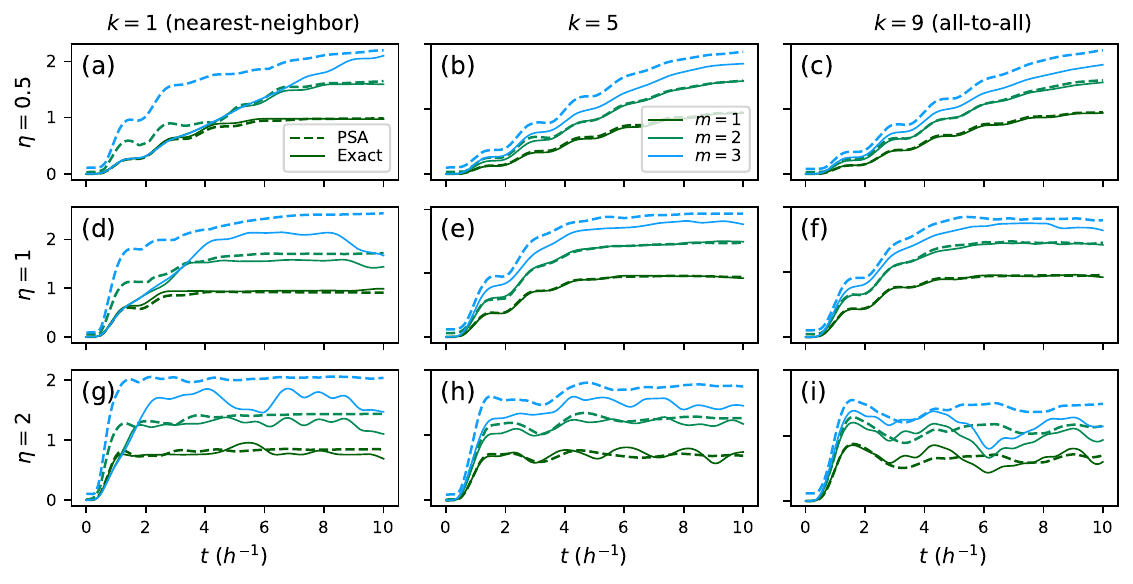}
    \caption{
        Comparison between the exact (solid lines) and PSA (dashed line) bipartite entanglement entropy between the leftmost $m = 1, \dots, 3$ qubits, obtained for $L = 10$ sites, $\eta \in \{ 0.5, 1, 2 \}$ and $k \in \{ 1, L/2, L-1\}$, for an evolution starting at the initial state $\ket{+}^L$.
    }
    \label{fig:multi-qubit-entropies}
\end{figure*}

Equation~\eqref{eq:multi-qubit-entanglement-entropy} requires the diagonalization of a $2^m \times 2^m$ (hermitian) matrix, so that practical computation is limited to small values of $m$. 
In Fig.~\ref{fig:multi-qubit-entropies}, we thus display the evolution of $S_I(t)$ for $I = (1, \dots, m)$ where $m \in \{ 1, 2, 3\}$, for $L=10$, obtained from both the PSA and the exact evolutions.
This corresponds to the entanglement entropy of the 1, 2, and 3 leftmost qubits with respect to the rest of the chain.
Note that we use the logarithm in base 2.

For all values of $k$ and $\eta$, we see from this figure that the PSA approach has a good predictive power for the 1-qubit entanglement entropy.
It is noteworthy that the 1- and 2-qubit entanglement entropies previously computed with the PSA for neutrinos with all-to-all interactions~\cite{Lacroix2022,Lacroix2024,Mangin-Brinet2025} were found to be in excellent agreement with the exact values. 

Here, however, the 2-qubit entanglement entropy is reproduced with a lesser fidelity, especially at $k = 1$ and at short times.
In this regime, we observe that the PSA tends to overestimate the entanglement entropy at short times, with an inaccurate initial growth rate.
This unwanted behavior worsens when the subsystem size increases, a phenomenon which may be attributed to small instabilities in the $2^m$ eigenvalues of $R_I$ that contribute to $S_I$.
Such instabilities appear in fact at $t = 0$ for any value of $k$, resulting in $S_I(t=0) > 0$, and are due to statistical fluctuations of the random variable $R_I(t=0)$.
Indeed, at the initial time and with the Rademacher sampling, the density matrices of individual trajectories are not positive; only the average over trajectories should recover positivity in the limit of infinite trajectories.
Since the exact state is a product state, some eigenvalues of the exact density matrix are zero.
Therefore, for a finite number of trajectories, we expect statistical fluctuations to yield positive and negative eigenvalues at a distance $\sim 1 / \sqrt{N_{\rm traj}}$ from zero.

Finally, we mention that PSA does not reproduce the symmetry of entanglement entropy when swapping the two subsystems, i.e., if $I$ and $I'$ are two subsystems that partition the whole system, then $S_I = S_{I'}$.
This is a consequence of the fact that the PSA "state" $\overline{R}$ is a mixed state, while the mentioned symmetry holds only for pure states.
Actually, it is known that the von Neumann entanglement entropy Eq.~\eqref{eq:multi-qubit-entanglement-entropy} is not a good indicator of entanglement in mixed states\cite{bennett_mixed-state_1996, wootters_entanglement_1998}.
For all these reasons, we restrict ourselves to the study of 1-qubit reduced entropies in the next section.

\section{Benchmark on large-scale TFIM}
\label{sec:psa-benchmark-large-scale}

One very attractive aspect of the PSA approach is its quadratic numerical scaling and therefore its capacity to simulate very large numbers of qubits, in any dimension. 
These aspects are studied below.
Based on the conclusions above, we focus on one-qubit observables and moderate coupling strengths $\eta < 2$ for which the PSA is expected to be accurate.

\subsection{Numerical scaling for large numbers of qubits}

\begin{figure}
    \centering
    \includegraphics[width=1\linewidth]{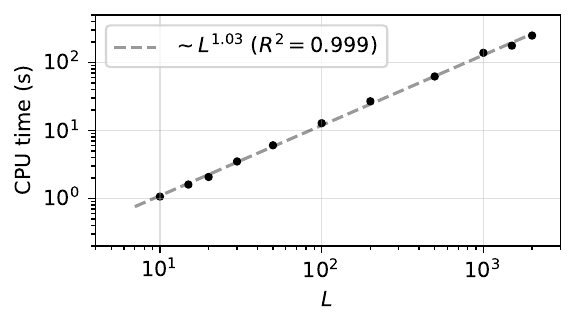}
    \caption{
    Linear scaling of the CPU time required for the PSA evolution of chains with varying length $L$ and nearest-neighbor coupling, using $N_\textrm{traj} = 500$ trajectories and 200 processors, including pre-computation of $3L$ commutators with $H$.
    Dashed grey line: linear regression performed over the data points.
    }
    \label{fig:num-scaling-duration}
\end{figure}

As the number of qubits $L$ increases, the number of coupled equations to solve is always $3L$.
Therefore, we expect them to be solved with a low polynomial complexity.
For instance, we performed their evolution in our work with a non-adaptive 4-th order Runge-Kutta method to evolve all PSA trajectories.
Such solver typically incurs a linear cost with respect to the number of equations.
As a result, we confirm an overall linear scaling in Fig.~\ref{fig:num-scaling-duration}, where we show the CPU time required to compute $N_\textrm{traj} = 500$ PSA trajectories of a 1D TFIM with $k=1$ over $10^3$ time steps and with 200 processors.

In general, we expect the overall complexity to go from $\mathcal{O}(L)$ when $k = 1$ up to $\mathcal{O}(L^2)$ when $k \sim L$.
This drastic increase is due to the computation of all $3L$ commutators from Ehrenfest's equations insofar as each single commutator requires itself $\mathcal{O}(L)$ terms (Pauli strings) to be built in the limit of all-to-all coupling.
Thus we regard the complexity of our approach as being quadratic, although in many prominent cases (e.g. nearest-neighbor or second-nearest-neighbor coupling) it remains linear in $L$.
Finally, we emphasize that one can always benefit from the independence of the different trajectories, which allows to parallelize them on several CPUs.

\subsection{Single qubit observables}

On small systems, as we saw in the previous sections, single-qubit observables are well described by PSA.
Here we extend this claim to large systems, up to $L=2000$ qubits.
Since exact computation is out of reach at these scales, we compare PSA with a Matrix Product State (MPS) calculation.
The quality of the MPS approximation is tuned by its bond dimension $\chi$, which characterizes the amount of entanglement entropy it captures~\cite{Orus2014}.
For a chain of length $L$, we restrict this bond dimension to $\chi = 2^9/\sqrt{L}$, a choice justified in App.~\ref{app:bond_dim_mps} by comparison to the memory cost of an exact resolution.
The bond dimension of the MPS decreases with $L$, so that the memory cost is caped at the memory used for an exact resolution with $\sim 20$ qubits. 

\begin{figure}
    \centering
    \includegraphics[width=1\linewidth]{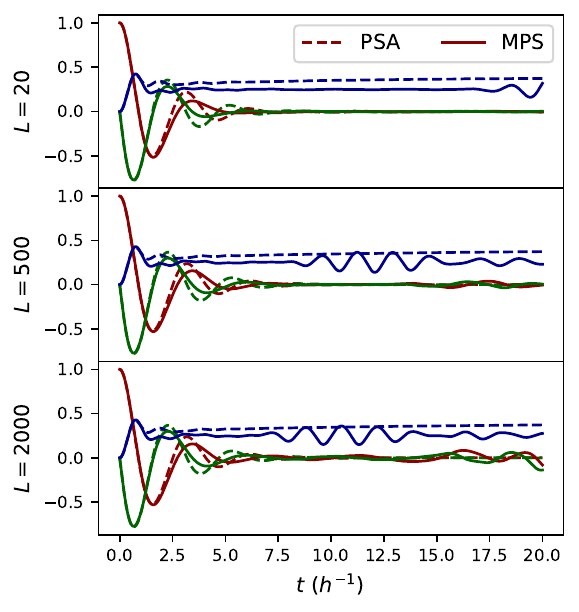}
    \caption{
    PSA and MPS evolution of the mean Bloch coordinates $\hat{S}_O$, $O = X, Y, Z$ (red, green, blue) in three chains of size $L = 20$, $500$ and $2000$ qubits, with $\eta = k = 1$ and initial state $\ket{+}^L$.
    }
    \label{fig:evol-long-chains}
\end{figure}

The comparison is shown in Fig.~\ref{fig:evol-long-chains}, where the evolution of average (over qubits) Bloch coordinates are displayed for increasing $L$.
Here, we observe a good agreement between the MPS and the PSA, accrediting the hypothesis that even at such large number of qubits, both methods yield results close to the exact physics of the TFIM.
In each panel, we observe an equilibration of the system, characterized by the convergence of the observables around an asymptotic mean value.
The results are similar when varying $L$; this can be explained by the emergence of a translational periodicity in the bulk of the chain, a phenomenon expected from the low variances between different qubits as mentioned in App.~\ref{app:inter-qubit-variance}.

Eventually, the main difference between both predictions proves to be small oscillations of $\expval{S_Z}$, located around $t = 10$ ($h^{-1}$) for $L \ge 500$ with the MPS.
By extrapolation from small length chains, we expect these oscillations to be inherent to the physics of the chain, rather than a numerical artifact from the method.

\begin{figure}
    \centering
    \includegraphics[width=1\linewidth]{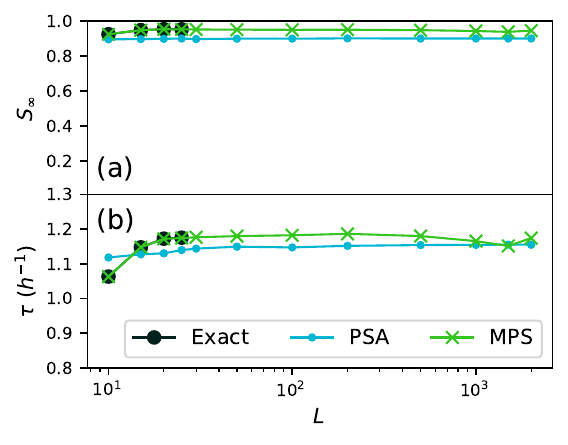}
    \caption{
    Dependence of the entropy asymptotic value $S_{\infty} \in [0, 1]$ (panel a) and characteristic growth time $\tau$ (panel b) with respect to $L$.
    Obtained from exact (black dots), PSA (blue circles) and MPS (green crosses) evolutions, with physical parameters: $\eta = k = 1$, initial state $\ket{+}^L$, evolution duration $T = 20$ ($h^{-1}$).
    }
    \label{fig:entropy-constants-long-chains}
\end{figure}

\subsection{Capturing equilibration processes}

\subsubsection{Key figures of equilibration}

We turn to an aspect of dynamics for which PSA is believed to provide particularly good estimates: the equilibration process.
Our goal here is to characterize the equilibration process of one-qubit degrees of freedom in the $k$-local TFIM described by Eq.~\eqref{eq:hamTFIM}. 
We focus on the possibility of extending our exact analysis of small chains ($L \le 20-30$) to larger systems using both the MPS, and the PSA.

One of the signatures of the equilibration process is the damping observed in the Bloch coordinates, as previously shown in Fig.~\ref{fig:exa-psa-mf-comparison}, panel (e).
Examples of dampened Bloch coordinates are given in Fig.~\ref{fig:evol-long-chains}.

Another signature, which is more convenient, is the stabilization of the 1-qubit entanglement entropy over time, observed in Fig.~\ref{fig:multi-qubit-entropies}.
Here, we will use the average (over qubits) 1-qubit entanglement entropy 
\begin{equation}
    S
    =
    -{1 \over L}
    \sum_{i = 1}^L
    \Tr[\overline{R_i}\log \overline{R_i}].
\end{equation}
Interestingly, this quantity depends only on the norm of the Bloch vectors $\Vert \overline{\vec{r}_i} \Vert$.
Indeed, introducing $\lambda_i^\pm = (1 \pm \Vert \overline{\vec{r}_i} \Vert)/2$ the two eigenvalues of $\overline{R}_i$, we have
\begin{equation}
    \Tr[\overline{R_i}\log \overline{R_i}]
    =
    \lambda_i^+\log(\lambda_i^+)
    +
    \lambda_i^-\log(\lambda_i^-),
\end{equation}
Hence, the 1-qubit entropy is a convenient way to summarize information about the dampening of Bloch coordinates.
As can be seen in Fig.~\ref{fig:multi-qubit-entropies}, we expect for most parameters that $S(t)$ first increases rapidly during a characteristic time $\tau$ before saturating to an asymptotic value $S_\infty$.
We will focus on these two quantities for the remainder of this section, using the definitions that follow.

First, to account for small fluctuations of $S$, we take the asymptotic entropy to be
\begin{equation}
    S_\infty = {1 \over \Delta }\int_{T-\Delta T}^{T}S(t) \textrm{d}t. 
    \label{eq:s-asymp}
\end{equation}
Here we used $T= 20 \; (h^{-1})$, while $\Delta T = 4 \; (h^{-1})$, in order to smooth out small fluctuations of $S$.
Second, the characteristic time $\tau$ is defined by fitting the entropy over an exponential ramp of the form $t \longmapsto S_\infty (1 - \exp -t/\tau)$.
Note that we could also determine $S_\infty$ through this fit, but in practice, this worsens the sensitivity of $\tau$ to accidental quirks in the entropy due, e.g., to finite-size effects.

\subsubsection{Results}

We show these quantities for different length chains $L$, ranging from $L = 10$ to $L = 2000$ qubits in Fig.~\ref{fig:entropy-constants-long-chains}.
Overall, we see that the PSA estimations of both the long-time entanglement entropy and the characteristic raising time are, although not equal, consistent with the MPS calculation.
Given that the two techniques are vastly different in nature, this comforts us in saying these estimations are close to exact.
The small difference between PSA and MPS can be explained by a small discrepancy in the asymptotic value of the Bloch coordinates, where PSA overestimates the slightly oscillating $Z$ coordinate, as shown in Fig.~\ref{fig:evol-long-chains}.

While these physical quantities are interesting in their own, the main takeaway is that the PSA appears in good agreement with other approximate methods like MPS.
Moreover, this illustration shows that PSA is suitable for very large systems, up to thousands of qubits, to capture important aspects of the dynamics. 

\begin{figure}
    \centering
    \includegraphics[width=1.0\linewidth]{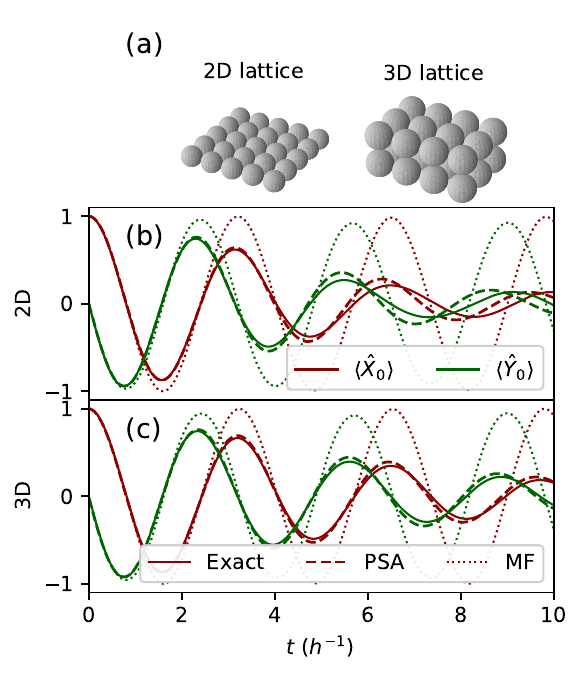}
    \caption{
    Comparison between PSA, MF, and exact dynamics in 2D and 3D.
    (a) Geometries used for 2D and 3D simulations.
    (b, c) Dynamics of $\langle X_0 \rangle$ and $\langle Y_0 \rangle$ on one of the corner qubits using the exact (solid lines), MF (dotted), and PSA (dashed) methods.
    $\expval{Z_0}$ is as well reproduced by the PSA as $\langle X_0 \rangle$ and $\langle Y_0 \rangle$, but is omitted here to improve the legibility.
     Parameters: initial state $\ket{+}^L$, $\eta = 0.75$ and $k = 1$.
    }
    \label{fig:2D3Dsimulations}
\end{figure}

\subsection{Generalization to 2D or 3D lattices}

The MF approach, as described in Sec .~\ref {sec:mfstrategy}, can be generalized to multi-dimensional lattices, with an arbitrary interaction matrix, and so does the PSA approach.
In this section, we compare PSA with MF and with exact dynamics on 2D and 3D square-lattice TFIM, with only nearest-neighbor interactions.
The specific lattices are shown in Fig.~\ref{fig:2D3Dsimulations}-(a). They correspond to a $5\times 5$ and a $4\times 3 \times 2$ lattices.
For such a number of qubits, the exact solution is still reachable.

Fig.~\ref{fig:2D3Dsimulations} (b) and (c) show the dynamics of the $X$ and $Y$ Bloch coordinates of one of the corner qubits, respectively for 2D and 3D.
In these examples, we clearly see that the MF does not capture any damping effect, whereas the PSA provides a much better estimate, only slightly underestimating the damping. 
Interestingly, when going from 2D to 3D, we observe a slight improvement in the PSA in the second half of the evolution, as the phase of the oscillations appears more accurate in 3D.
Despite the different geometries and boundary effects, this could be explained by the higher number of neighbors, which makes the MF approximations in the PSA less detrimental.

\section{Conclusion}

The Phase-Space Approximation, originally proposed to treat interacting fermionic systems, is formulated here directly on qubits and benchmarked on an Ising model. 
Based on a statistical sampling of independent mean-field evolutions, this technique has a low numerical cost, allowing simulations of qubit registers containing up to several thousand qubits. 

First, we provided a simple formulation of PSA on qubits, refocusing the attention on the density matrix and the 1-qubit observables, which are more adapted to the quantum computing community.
By equivalence, this also provides an alternate, and simple, formulation of the discrete truncated Wigner approximation.

Then, we benchmarked the approach across a diverse set of physical problems encoded on qubits. 
To this aim, we considered the TFIM with various levels of locality and interaction strength. 
We simulated systems ranging from nearest-neighbor to all-to-all interactions, which are respectively relevant to condensed matter and nuclear physics.

The original MF theory, from which the PSA approach is formulated, implicitly assumes that one-qubit degrees of freedom contain the important information on the system. 
As such, MF is adequate for short-time evolution and, in general, is unable to describe quantum fluctuations. 
While keeping the simplicity of MF along each trajectory, the PSA technique has a much better predictive power. 
We showed that, for most parameters, it provides better reproduction of one-qubit properties than MF and includes damping effects towards well-predicted long-time equilibrium. 
Although it also improves the description 
of two-qubit observables and/or quantum fluctuations compared to MF, this reproduction is not perfect.
We observe deviations in the prediction of observables, especially multi-qubit observables, and in the local interaction regime characterized by $k \ll L$.
These results complete previous observations made on fermionic systems in the case of all-to-all systems, such as atomic nuclei or neutrinos.

Our main conclusion from this study is that the PSA approach, although not exact, provides a good reference for comparison of 1-qubit properties for future simulation on quantum devices.  
This technique might be of great interest with large qubit registers, where other approximate methods might be extremely difficult to apply, or would benefit from a double-check when applicable (e.g. tensor networks).
As an illustration of the attractive aspects of the PSA methods, applications to 1D systems having up to 2000 qubits, and systems in 2D or 3D, are presented and compared to tensor network emulation or exact solutions when possible.

Hopefully, the PSA will add to state-of-the-art quantum simulation methods, thanks to its simplicity, computational efficiency, and availability for different geometries.
In particular, our formulation for qubits complements the formulation of the discrete truncated Wigner approximation, which yields similar equations through the Wigner-Weyl formalism~\cite{kunimi-2021-twa-spin-systems, khasseh-2020-dTWA-Ising, singh-2022-dTWA, shenoy-2024-dtwa-neural-qstates, hosseinabadi-2025-twa-dissipative}.
We leave for future work the theoretical development of the PSA based on refinements of the MF techniques, and further studies based on clustered geometries~\cite{wurtz-2018-cluster-twa, braemer-2024-cluster-twa}.

\section{Acknowledgments} 

We sincerely thank the referees, who took time to review our article, and provided us with relevant changes and constructive criticism.
This project has received financial support from the CNRS through the AIQI-IN2P3 project.
This work is part of the HQI initiative (\href{www.hqi.fr}{www.hqi.fr}) and is supported by France 2030 under the French National Research Agency award number ``ANR-22-PNQC-0002''.
This publication has received funding under Horizon Europe programme HORIZON-CL4-2022-QUANTUM-02-SGA via the project 101113690 (PASQuanS2.1).
Exact and MPS numerical calculations were performed using the Eviden Qaptiva platform.

%

\newpage
\appendix

\section{Discussion of exact versus approximate MPS simulations}

We discuss here the current frontier between exact and approximate simulation using the MPS technique of a qubit register. Such a frontier depends on the capabilities of current classical computers. We dissociate the use of MPS on personal computers from its use on High-Performance Computing (HPC) machines. A preliminary discussion is made first on the bond dimension, which is the key ingredient in MPS. 

\subsection{Choice of bond dimension and memory usage of MPS}
\label{app:bond_dim_mps}

The memory footprint of an MPS is characterized by the number of sites $L$, the dimension of the Hilbert space of a site $d$, and the bond dimension $\chi$.
If the MPS has varying bond dimensions along the sites, consider $\chi$ to be the maximal one.
Then, an upper bound of the memory usage of the MPS is
\begin{eqnarray}
    M
    =
    L d \chi^2, \label{eq:chimax}
\end{eqnarray}
in units of the number of complex values.
This upper bound ignores the fact that bond dimensions are lower close to the edges.
This is a good approximation if $\chi \ll d^{L/2}$, which in practice is always the case at large $L$.

As a result, the largest MPS that can fit a memory volume $M$ has a bond dimension $\chi_{\rm max} \approx \sqrt{M / Ld}$.
Two important cases, using $d=2$ (i.e. qubits):
\begin{enumerate}
    \item On a laptop, we can assume the memory can fit a state vector of about 19 qubits (A complex number in double precision is 16 bytes, so that a 19 qubits state vector represents about 10MB), i.e. $M=2^{19}$. This gives $\chi_{\rm max} = 2^9 / \sqrt{L}$.
    
    For $L = 64 = 2^{6}$ this gives $\chi_{\rm max} = 64$.
    
    For $L = 1024 = 2^{10}$ this gives $\chi_{\rm max} = 16$.
    
    \item On a large HPC machine, we can assume the memory can fit a state vector of about 37 qubits ( a 37 qubit state vector represents about 2TB), i.e., $M=2^{37}$. This gives $\chi_{\rm max} = 2^{18} / \sqrt{L}$.
    
    For $L = 1024 = 2^{10}$ this gives $\chi_{\rm max} \approx 8000$.
    
    For $L \approx 65000 \approx 2^{16}$ this gives $\chi_{\rm max} \approx 1000$.
\end{enumerate}

\subsection{$k$-local TFIM model exactly solved on classical computers}
\label{sec:mps}

For moderate $L$ values, direct diagonalization of the Hamiltonian allows for performing the exact dynamics of any initial state. 
Leaving for the moment the problem of the CPU time, such direct solutions in a $2^L$-dimensional Hilbert space are doable on a desk computer for $L \simeq 12-18$ and on a HPC machine for $L  \simeq 30-38$ (up to 41 or 42 for very large HPC). 
A critical aspect of the advances in quantum computers is the possibility of performing reliable, though perhaps approximate, classical computer simulations of a problem on future machines with several hundred or thousands of qubits.  

In some cases, tensor network techniques such as MPS might provide such a benchmark provided that the entanglement entropy between two subsystems inside the whole system remains small enough. As discussed 
In appendix \ref{app:bond_dim_mps}, the important parameter is the maximal bond dimension, denoted by $\chi_{\rm max}$, allowed by the memory $M$ of the used classical computer architecture.  Eq. (\ref{eq:chimax}) gives 
for a one-dimensional 
chain of length $L$: $\chi_{\rm max} \approx \sqrt{M / Ld}$ with $d=2$ for qubits.
In addition, we have the constraint~\cite{Ayral2023}: 
\begin{eqnarray}
2^{S_{\rm max}} \le \chi_{\max} \le 2^{L/2}.
\end{eqnarray}     
where $S_{\rm max}$ is the maximal entanglement entropy that can be reproduced with an MPS technique limited by $\chi_{\max}$. 
Summing up, assuming that $M=2^{m}$, we see that we have the condition 
of the maximal entropy that could be simulated using MPS:
\begin{eqnarray}
S_{\rm max} \le \min\left(\frac{L}{2} , \frac{(m-1)}{2} -\frac{1}{2} \log_2 (L)  \right).
\end{eqnarray}

\begin{figure}[]
\centering
\includegraphics[width=1\linewidth]{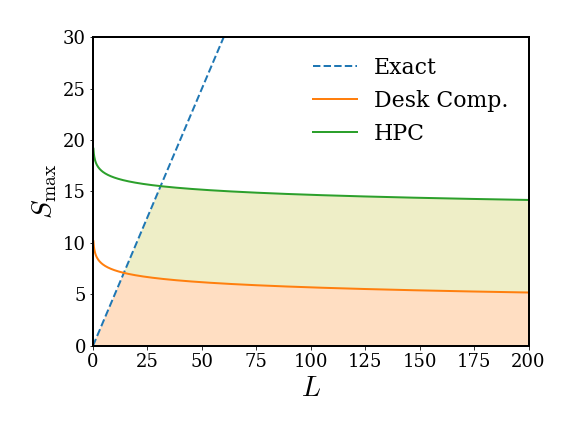}
    \caption{
    Illustration of the maximal entanglement entropy between two subsystems of a qubit register as a function of the number of qubits/sites $L$ that could be simulated on a classical computer using the MPS technique. 
    The dashed blue line corresponds to the absolute maximum in the exact solution.  
    The orange and green solid lines correspond to the curve 
    $(m-1)/2 -[\log_2 (L)]/2$ for desk computer ($m = 19$) and 
    HPC ($m = 39$) machines respectively (see appendix \ref{app:bond_dim_mps} for more details).
    The orange area indicates the region of $(L, S_{\rm max})$ that could be efficiently simulated on a desk computer based on the MPS framework. The green 
    area indicates the extension of the region using HPC machines.}
    \label{fig:entmax}
\end{figure}
A schematic view of the maximal entanglement entropy that could be simulated using MPS is shown in Fig. \ref{fig:entmax}. 
As shown in Refs.~\cite{Cal05,Alb18}, for specific initial conditions, the entanglement entropy of an equal bi-partition of the system in the $1$-TFIM presents a ballistic evolution leading to an asymptotic 
entropy that scales linearly with $L$ at short times.   
As extensively discussed in Ref.~\cite{Ans24}, this constitutes a stringent challenge for the MPS theory.
In addition, it was shown that the account of long-range entanglement 
induced, for instance, in the $2$-local TFIM model with an MPS 
requires specific attention \cite{Frias2024}.

\section{Additional results for 1-qubit observables}

\subsection{Influence of the initial state}

\label{app:influence-initial-state}

In this section, we detail the protocol used to study the effect of the initial state $\ket{\theta, \phi}$ on the error, quantified through the difference $D_r(T)$ for both the MF and PSA.

Numerically, we sample coordinates $(\theta, \phi)$ which are respectively the colatitude and the longitude on the Bloch sphere.
This corresponds to the convention where the states $\ket{0}$ and $\ket{1}$ are the North and South poles, with latitude $\pm \pi/2$.
The state $\ket{+}$ investigated in a part of this article has zero longitude and latitude.
We perform the sampling of all angles using a Fibonacci sampling to ensure an even spacing~\cite{gonzalez-2010-fibonacci-sampling}.

Graphically, we represent the sampled values of $D_r(T)$ through an irregular heatmap.
Its cells are defined by the Voronoi tessellation induced by the sampled points.
The Voronoi tessellation is computed directly on the 3D Bloch sphere to avoid boundary effects near the antimeridian (longitude $\pm \pi$).
The results are mapped to a 2D Mollweide projection of the Bloch sphere; this ensures area preservation between the final result in Fig.~\ref{fig:psa-heatmap-mollweide} and the cells lying on the 3D Bloch sphere~\cite{snyder-1987-map-projections}.

\begin{figure}[ht]
     \centering
     \includegraphics[width=1\linewidth]{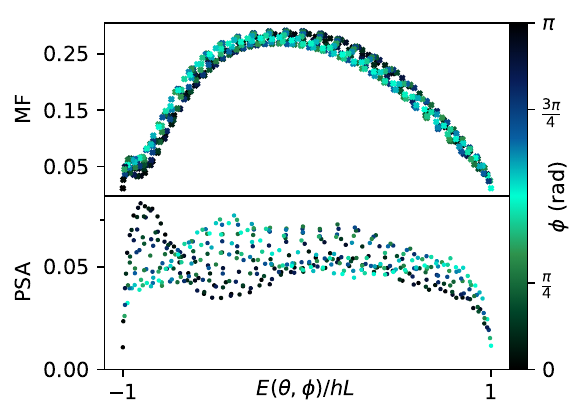}
     \caption{
     Dependence of the deviation $D_r(T)$ in the MF energy $E(\theta, \phi)/hL$ of the corresponding initial state, using the parameters of Fig.~\ref{fig:psa-heatmap-mollweide}.
     }
     \label{fig:mollweide-correlations}
 \end{figure}

In addition to the main text, we show in Fig.~\ref{fig:mollweide-correlations} how the discrepancy $D_r(T)$ correlates with the MF energy $E(\theta, \phi)$ given by Eq.~\eqref{eq:mf-ansatz-energy} for a given initial state $\ket{\theta, \phi}^L$.
We see again that the PSA error is lower than MF on the whole Bloch sphere. Only near the poles ($\ket 0$ and $\ket 1$) does the MF error get close to, but still higher than, PSA error.

\subsection{Systematic results along a grid of parameters}
\label{app:systematic-scanning}

\begin{figure*}[!htbp]
    \centering
    \includegraphics[width=1\linewidth]{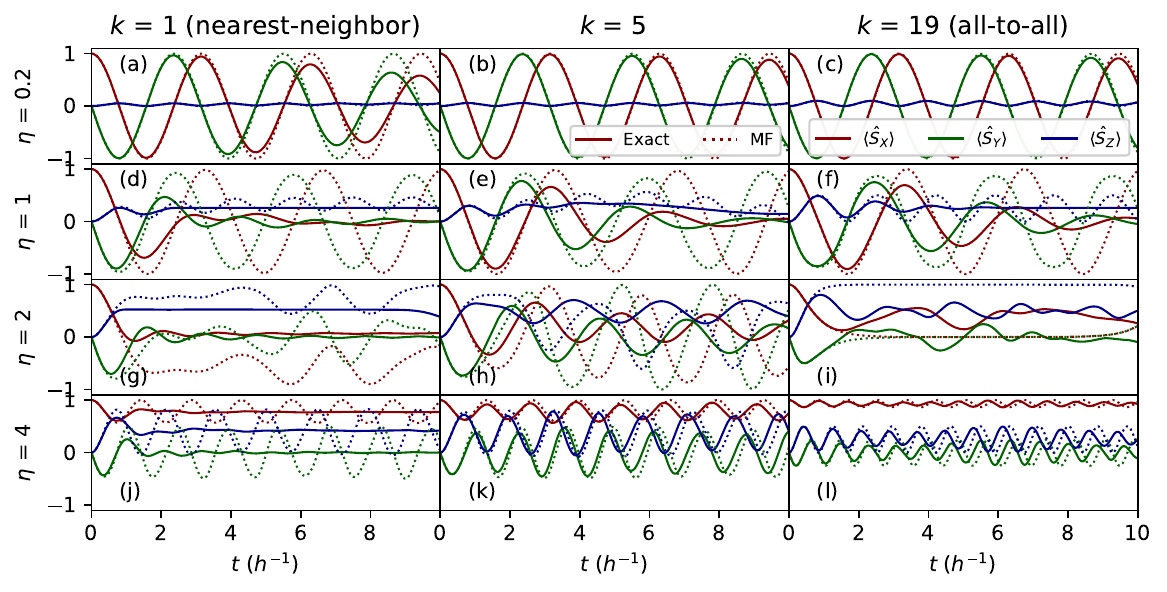}
    \caption{
    Comparison of the exact (solid line) and MF (dotted line) evolutions of Pauli matrices expectation values, with the same protocol as in Fig.~\ref{fig:exa-psa-mf-comparison}.
    }
    \label{fig:full-exa-mf-comparison}
\end{figure*}

\begin{figure*}[!ht]
     \centering
     \includegraphics[width=1\linewidth]{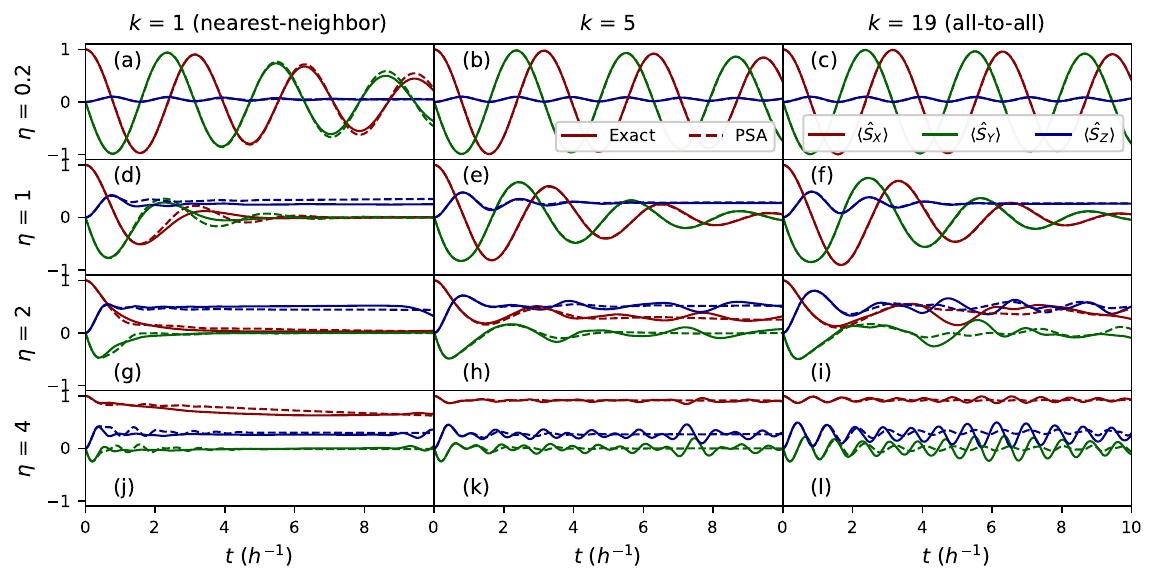}
     \caption{
     Comparison of the exact (solid lines) and PSA (dashed lines) dynamics, obtained by averaging $N_\textrm{traj} = 10^{4}$ trajectories.
     Same observables and same parameters as in Fig.~\ref{fig:full-exa-mf-comparison}, with a comparison of both figures showing a drastic improvements thanks to the PSA.
     }
     \label{fig:full-exa-psa-comparison-obs-avg}
 \end{figure*}

An important part of our MF-PSA comparison is displayed in Fig.~\ref{fig:exa-psa-mf-comparison} for a few pairs of ($\eta$, $k$) parameters.
These pairs were chosen to illustrate the three most prominent physical regimes observed when varying the two main parameters, for a given initial state $\ket{+}^L$.
For the sake of completeness, we provide in Fig.~\ref{fig:full-exa-mf-comparison} and Fig.~\ref{fig:full-exa-psa-comparison-obs-avg} the MF and PSA results for a broader scanning of the parameter space.

Notably, we observe in the strong coupling regime ($\eta \ge 2$) that the long-time revival of oscillations are not properly reproduced by the PSA; see, e.g. panel (k) of Fig.~\ref{fig:full-exa-psa-comparison-obs-avg}.
This revival is due to quantum interferences induced by the bouncing back of the wave function at the boundaries of the lattice.
Such effects cannot be treated assuming independent classical trajectories; we refer the interested reader to the related discussions in Refs.~\cite{Regnier2018, Regnier2019}.

\subsection{Variation of single-qubit observables across qubits}
\label{app:inter-qubit-variance}

\begin{figure*}[!ht]
     \centering
     \includegraphics[width=1\linewidth]{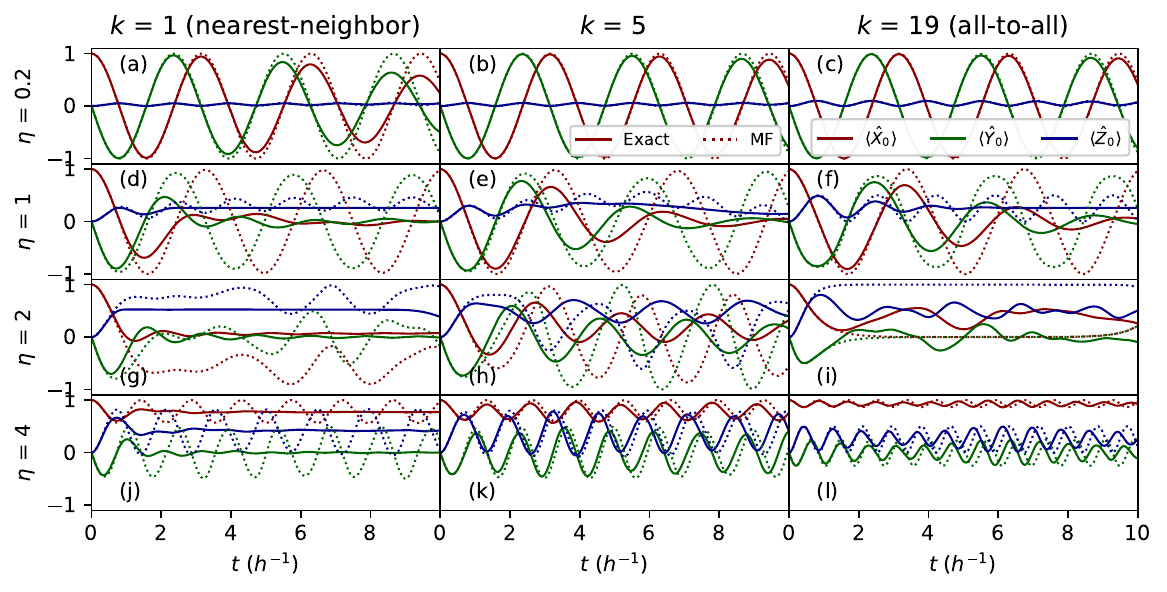}
     \caption{
     Equivalent of Fig.~\ref{fig:full-exa-mf-comparison} with the Bloch coordinates of the first qubit of the chain instead of an average over all $L$ qubits.
     }
     \label{fig:exa-mf-comparison-qubit0}
 \end{figure*}

\begin{figure*}[!ht]
     \centering
     \includegraphics[width=1\linewidth]{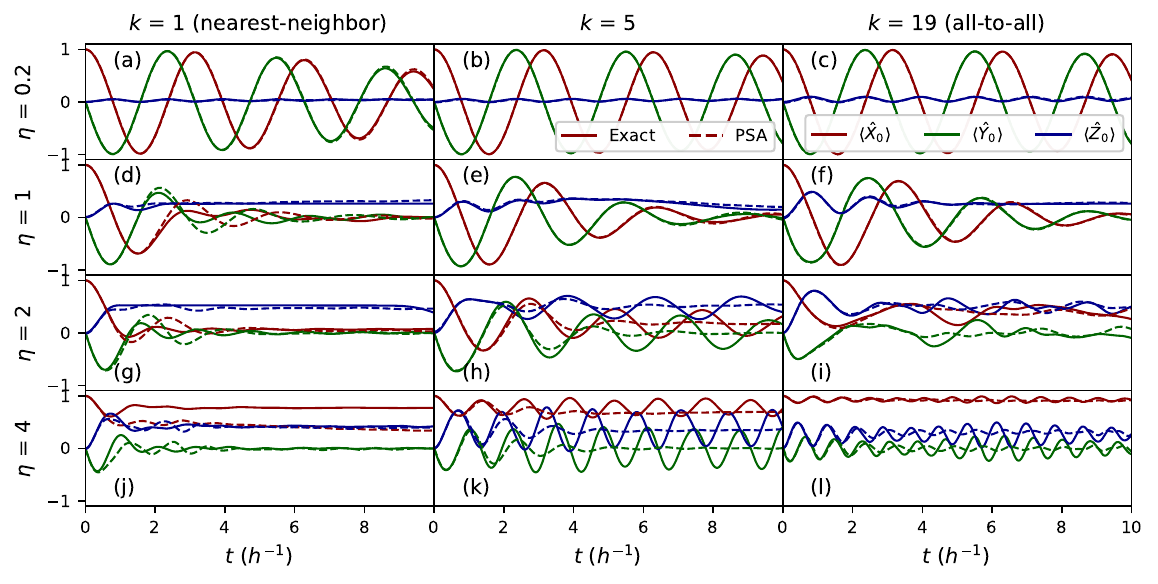}
     \caption{
     Equivalent of Fig.~\ref{fig:full-exa-psa-comparison-obs-avg} with the Bloch coordinates of the first qubit of the chain instead of an average over all $L$ qubits, with visible differences in panels (g, h, j, k).
     }
     \label{fig:exa-psa-comparison-qubit0}
 \end{figure*}

We show in Figs.~\ref{fig:exa-mf-comparison-qubit0}-\ref{fig:exa-psa-comparison-qubit0} the time-dependent Bloch coordinates of the first qubit of the chain (i.e., the qubit at the end of the chain) evolved with the same parameters as in Fig.~\ref{fig:full-exa-mf-comparison}-\ref{fig:full-exa-psa-comparison-obs-avg}.

Both the exact and PSA data are very close to their qubit-averaged counterpart, except in the range of parameters $k \le 5$, $\eta \ge 2$ (panels g, h, j, and k).
Moreover, we found that the standard deviation of any Bloch coordinate between the $L$ qubits does not exceeds 0.061 for the exact resolution, 0.0592 for the PSA, and 0.189 for the MF.
From these observations, we deduce that considering qubit-averaged observables is sufficient to get a good understanding of 1-qubit observables, in the scope of our benchmark.

\end{document}